\documentclass{emulateapj}

\usepackage{rotating}
\usepackage{natbib}
\usepackage{apjfonts}
\newcommand\tabspace{\noalign{\vspace*{0.7mm}}}
\def\errtwo#1#2#3{$#1^{+#2}_{-#3}$}
\def\errpm#1#2{$#1^{+#2}_{-#2}$}
\newcommand\tfx{{F_{\rm x}}}
\newcommand\tfr{{F_{\rm r}}}

\newcommand\cfr{{\cal F}_{\rm r}}
\newcommand\gx{{GX~339$-$4}}
\newcommand\cyg{{Cyg~X-1}}
\newcommand\isis{{\tt ISIS}}
\newcommand\atca{\textsl{ATCA}}

\newcommand\rxte{\textsl{RXTE}}
\newcommand\pca{\textsl{PCA}}
\newcommand\asm{\textsl{ASM}}
\newcommand\hexte{\textsl{HEXTE}}

\newcommand\aproxgt{\mathrel{%
      \rlap{\raise 0.511ex \hbox{$>$}}{\lower 0.511ex \hbox{$\sim$}}}}
\newcommand\aproxlt{\mathrel{%
      \rlap{\raise 0.511ex \hbox{$<$}}{\lower 0.511ex \hbox{$\sim$}}}}

\slugcomment{Submitted 2004, September 24}

\shorttitle{Is the `IR Coincidence' Just That?}
\shortauthors{Nowak et al.}

\begin{document}

\title{Is the `IR Coincidence' Just That?}

\author{Michael A. Nowak\altaffilmark{1}, J\"orn
Wilms\altaffilmark{2}, Sebastian Heinz\altaffilmark{1}, Guy
Pooley\altaffilmark{3}, Katja Pottschmidt\altaffilmark{4,5}, Stephane
Corbel\altaffilmark{5}} 
\altaffiltext{1}{Massachusetts Institute of Technology,
Center for Space Research, Cambridge, MA 02139, USA; mnowak,heinzs@space.mit.edu}
\altaffiltext{2}{Dept. of Physics, University of Warwick, Coventry, CV4 7AL, UK; 
wilms@astro.uni-tuebingen.de}
\altaffiltext{3}{Astrophysics Group, Cavendish Laboratory, Madingley Road, Cambridge 
CB3 OHE, UK; guy@mrao.cam.ac.uk}
\altaffiltext{4}{Max-Planck-Institut f\"ur extraterrestrische Physik, 
Giessenbachstra\ss{}e 85748 Garching, Germany}
\altaffiltext{5}{INTEGRAL Science Data Center, Chemin d'\'Ecogia 16, 1290 Versoix, 
Switzerland; Katja.Pottschmidt@obs.unige.ch}
\altaffiltext{6}{Universit\'e Paris VII and Service d'Astrophysique, CEA-Saclay, 
91191 Gif-sur-Yvette Cedex, France; corbel@discovery.saclay.cea.fr}

\begin{abstract}
Previous work by \citet{motch:85a} suggested that in the low/hard
state of \gx\ the soft X-ray power-law extrapolated backward in energy
agrees with the IR flux level.  \citet{corbel:02a} later showed that
the typical hard state radio power-law extrapolated forward in energy
meets the backward extrapolated X-ray power-law at an IR spectral
break, which was explicitly observed twice in \gx.  This `IR
coincidence' has been cited as further evidence that synchrotron
radiation from a jet might make a significant contribution to the
observed X-rays in hard state black hole systems.  We quantitatively
explore this hypothesis with a series of simultaneous radio/X-ray
observations of \gx, taken during its 1997, 1999, and 2002 hard
states.  We fit these spectra, in detector space, with a simple, but
remarkably successful, doubly broken power-law model that indeed
requires an IR spectral break.  For these observations, the break
position and the integrated radio/IR flux have stronger
dependences upon the X-ray flux than the simplest jet model
predictions. If one allows for a softening of the X-ray power law with
increasing flux, then the jet model can agree with the observed
correlation. We also find evidence that the radio flux/X-ray
flux correlation previously observed in the 1997 and 1999 \gx\ hard
states shows a `parallel track' for the 2002 hard state.  The
slope of the 2002 correlation is consistent with observations
taken in prior hard states; however, the radio amplitude is
reduced. We then examine the radio flux/X-ray flux correlation in
\cyg\ through the use of 15 GHz radio data, obtained with the Ryle
radio telescope, and \textsl{Rossi X-ray Timing Explorer} data, from
the \textsl{All Sky Monitor} and pointed observations.  We again find
evidence of `parallel tracks', and here they are associated with
`failed transitions' to, or the beginning of a transition to, the soft
state.  We also find that for \cyg\ the radio flux is more
fundamentally correlated with the hard, rather than the soft, X-ray
flux.
\end{abstract}

\keywords{accretion, accretion disks -- black hole physics --
radiation mechanisms:non-thermal -- X-rays:binaries}

\section{Introduction}\label{sec:intro}

\setcounter{footnote}{0}

Both \cyg\ and \gx\ in their spectrally hard, radio-loud states have
served as canonical examples of the so-called `low state' (or `hard
state') of galactic black hole candidates \citep[see][and references
therein]{pottschmidt:02a,nowak:02a}.  As we discuss below, in this
state the X-ray spectrum is reasonably well-approximated by a
power-law with photon spectral index (i.e., photon flux per unity
energy $\propto E^{-\Gamma}$) of $\Gamma \approx 1.7$, with the
power-law being exponentially cutoff at high energies ($\approx
100$\,keV).  It long has been suggested that such spectra are due to
Comptonization of soft photons from an accretion disk by a hot corona
in the central regions of the compact object system
\citep[e.g.,][]{sunyaev:79a}.  Comptonization models have been very
successful in describing the broad-band X-ray/soft gamma-ray spectra of
both \cyg\ \citep{pottschmidt:03a} and \gx\ \citep{nowak:02a}.

\begin{figure*}
\epsscale{1}
\plottwo{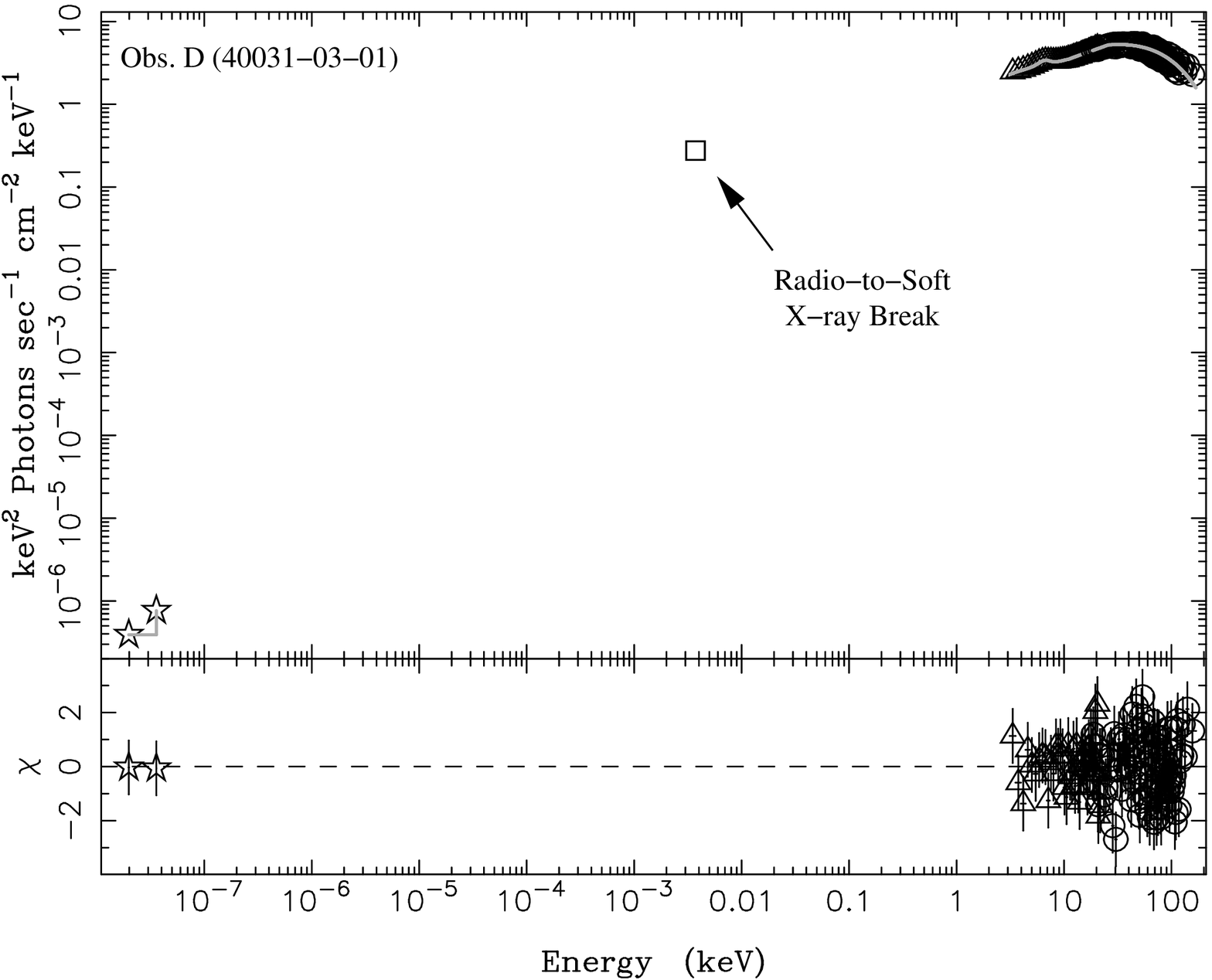}{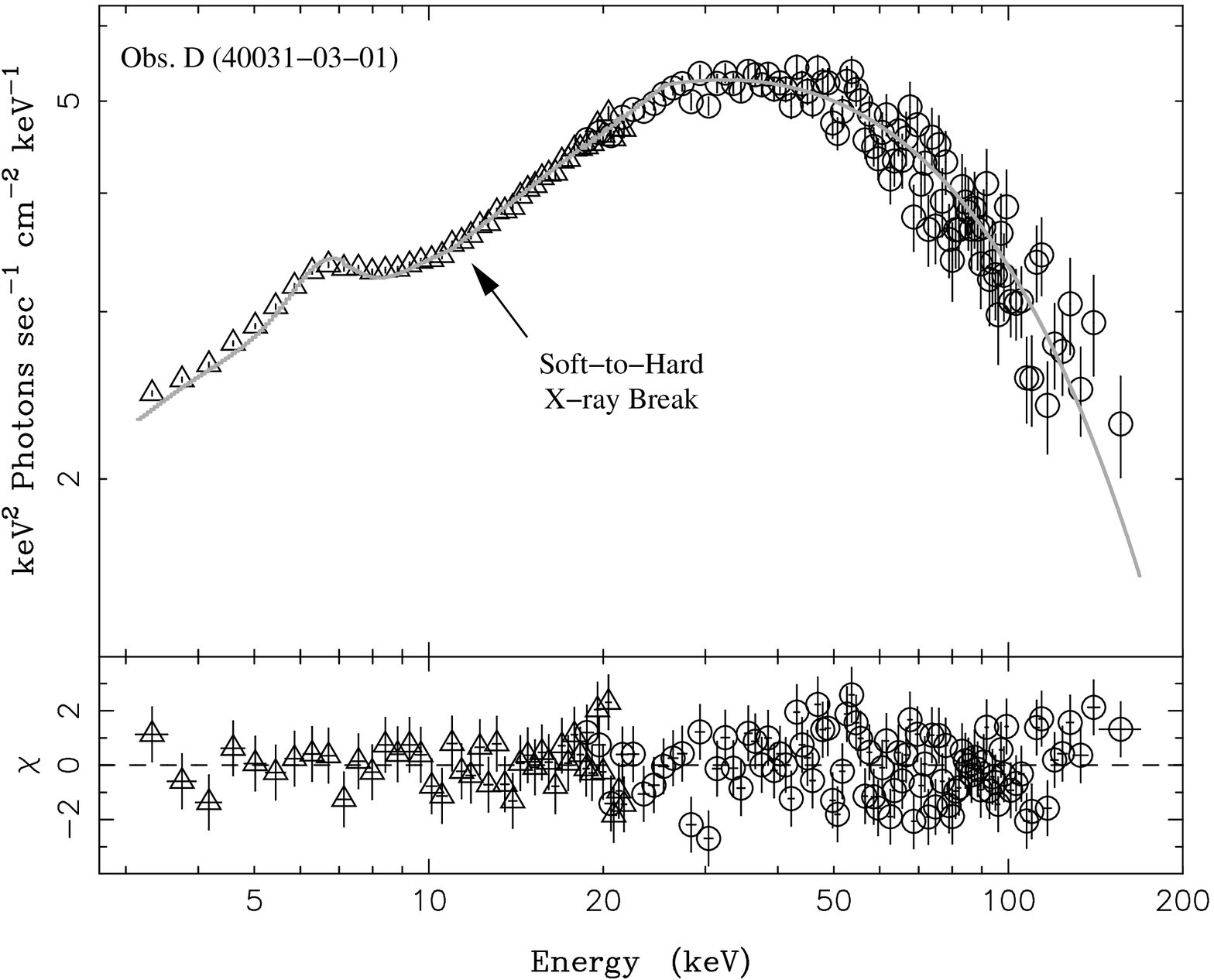}
\caption{Left: Unfolded spectra of a simultaneous radio and X-ray
spectrum of \gx\ (D, Observation ID 40031-03-01, in Tables 1 and 2),
fit with an absorbed, exponentially cutoff, doubly broken power-law
and a gaussian line. Residuals are from the proper forward folded
model fit.  Right: Same unfolded spectrum as on the left, showing just
the \rxte\ data. \label{fig:spectra}}
\end{figure*}  

It recently has been hypothesized, however, that the X-ray spectra of
hard state sources might instead be due to synchrotron and synchrotron
self-Compton (SSC) radiation from a mildly relativistic jet
\citep{markoff:01a,markoff:03a}.  Jet models have been prompted in
part by multi-wavelength (radio, optical, X-ray) observations of hard
state systems.  Although radio emission from \cyg\ was first observed
quite some time ago \citep{braes:71a}, it is more recently that a
radio jet has been imaged \citep{stirling:01a}, and that the low/hard
state X-ray flux has been shown to be positively correlated with the
radio flux \citep{pooley:98a}.

Radio emission also has been discovered in \gx\ \citep{sood:94a}.  The
radio emission is correlated with X-ray flux in spectrally hard states
\citep{hanni:98a}, but is quenched during spectrally soft states
\citep{fender:99b}.  Furthermore, in hard states of \gx, the 3--9\,keV
X-ray flux (in units of $10^{-10}~{\rm erg~cm^{-2}~s^{-1}}$) is
related to the 8.6\,GHz radio flux (in mJy) by $\tfx \approx 0.46
\cfr^{1.42}$ \citep[][noting that throughout this paper we shall use
caligraphic script to denote flux densities, i.e., flux per unit
energy, and roman script to denote flux integrated over an energy
band]{corbel:03a}.  This correlation was seen to hold over several
decades in X-ray flux, and also to hold for two hard state epochs that
were separated by a prolonged, intervening soft state outburst.
Similar correlations were found between energy bands in the 9--200\,keV
range and the radio flux \citep{corbel:03a}.  It further has been
suggested that the $\tfx \propto \cfr^{1.4}$ correlation is a
universal property of the low/hard state of black hole binaries
\citep{gallo:03a}.

This specific power-law dependence of the radio flux upon the X-ray
flux naturally arises in synchrotron jet models
\citep{falcke:95a,corbel:03a,markoff:03a,heinz:03a}. Both the location
of the break from an optically thick\footnote{By optically thick, we
mean radio energy flux density $S_\nu \propto \nu^{\alpha_{\rm r}}$,
with $\alpha_{\rm r} > 0$. Throughout this work, for both radio and
X-ray spectra, we will follow the convention $\alpha = 1 - \Gamma$.}
to an optically thin radio spectrum (presumed to continue all the way
through the X-ray), and the amplitude of the optically thick portion
of the radio spectrum, vary with input power to the jet so as to
produce the $\tfx \propto \cfr^{1.4}$ scaling.  Assuming, however,
that the X-ray power of a disk/corona is proportional to $\dot M^2$,
where $\dot M$ is the accretion rate, while the jet power is
proportional to $\dot M$, also reproduces the scaling relationship
\citep{markoff:03a,heinz:03a,fender:03a,merloni:03a,falcke:04a}.

Interestingly, nearly 20 years ago \citet{motch:85a} noted that for a
set of simultaneous IR, optical, and X-ray observations of the \gx\
hard state, the extrapolation of the X-ray power-law to low energy
agreed with the overall flux level of the optical/IR data.
\citet{corbel:02a} reanalyzed these observations, which did not
include simultaneous radio data, as well as a set of (not strictly
simultaneous) radio/IR/X-ray observations from the 1997 \gx\ hard
state.  They showed that the low energy extrapolation of the X-ray
power-laws, and the high energy extrapolation of the radio power-law,
coincided with a spectral break in the IR.  The spectral shapes below
and above the IR break were roughly consistent with the radio and
X-ray power-laws, respectively.  This `IR coincidence' has been cited
as further evidence that the jet not only accounts for the flat
spectrum from radio through infrared, but also that optically thin
emission from the jet, occurring at energies above the IR break,
provides a significant contribution to the observed X-rays
\citep{corbel:02a,corbel:03a,markoff:03a}.

In this paper, we quantitatively examine the `IR coincidence' with a
series of simultaneous radio/X-ray spectra of \gx.  We use broken
power-law fits to the combined radio and X-ray spectra to determine
the extrapolated position of the break as a function of observed X-ray
flux.  We then reassess the correlation between radio and X-ray flux
for \gx\ by including more recent hard state observations that
occurred at fairly high X-ray fluxes.  Based upon our results for \gx,
we also reassess this correlation for the case of \cyg.  We then
summarize our results.

\section{The `IR Coincidence' in \gx}\label{sec:ircoin}
\subsection{Data Analysis}\label{sec:anal}

We consider a set of ten simultaneous radio/X-ray observations of \gx,
eight of which were discussed previously by us
\citep{wilms:99aa,nowak:02a,corbel:03a} and come from the 1997 or 1999
hard state, and two of which were discussed by \citet{homan:04a} and
come from the 2002 hard state (approximately a month before a soft
state transition).  All X-ray observations were performed with the
\textsl{Rossi X-ray Timing Explorer} (\rxte). Their observation IDs
and associated radio flux densities and integrated X-ray fluxes are
presented in Table~1.  Note that four of these observations are
further labeled A--D, as we single these out for special discussion
below. Observations A and B (40108-01-03 and 40108-01-04) occurred
immediately after the 1999 soft-to-hard state transition
\citep{nowak:02a} and have optically thin radio spectra ($\alpha_{\rm
r} < 0$).  Observation C and D (70109-01-02 and 40031-02-01) have the
brightest X-ray fluxes in our sample, and are among the brightest hard
X-ray states observed in \gx\ to date.

To analyze the X-ray spectra of these observations, \rxte\ response
matrices were created using the software tools available in {\tt
HEASOFT 5.3}\footnote{The use of {\tt HEASOFT 5.3} is very important
here, as we find extremely good agreement between the
\textsl{Proportional Counter Array} and \textsl{High Energy X-ray
Timing Explorer} when fitting power-law models to the Crab pulsar plus
nebula system.  This is true for both the power-law normalization and
slope, both of which must be determined very accurately when
extrapolating over a large range of energies between the radio and
X-ray spectra.  This spectral and flux agreement is in marked contrast
to earlier versions of {\tt HEASOFT} \citep[see, for example, the
discussion of][]{wilms:99aa}.}.  The \textsl{Proportional Counter
Array} (\pca) data were rebinned to have a minimum of 30 counts per
bin, uniform systematic uncertainties of 0.5\% were applied, and only data
between 3 and 22\,keV were considered.  The \textsl{High Energy X-ray
Timing Explorer} (\hexte) data were coadded from the two individual
clusters and then were rebinned to have a minimum signal-to-noise
ratio (after background subtraction) of 10 in each bin.  We considered
\hexte\ data only in the 18--200\,keV range.  In the fits discussed
below, a multiplicative constant was allowed between the \pca\ and
\hexte\ normalization, with the \pca\ constant fixed to unity. The
normalizations of the two instruments were always found to be the same
to within a few percent (see Table~2).

The radio data for observation D (40031-03-01) were obtained with the
\textsl{Australia Telescope Compact Array} (\atca) at 4.8\,GHz and
8.6\,GHz.  The radio data for observation C (70109-01-02) were also
obtained with \atca, but at 1.4 and 2.4\,GHz.  These radio data, as
well as the data for observation B (40108-01-04), the 1999 April 22
observation (40108-02-02), and the 1999 May 14 observation
(40108-02-03) have been reanalyzed by us.  All other radio data come
from \citet{wilms:99aa}, \citet{nowak:02a}, and \citet{corbel:00a}.
In cases of small discrepancies between these latter references, we
have adopted the radio flux values of \citet{corbel:00a}, since, aside
from the radio observations reanalyzed by us, that work represents the
most up to date analysis of the radio data presented here.

Note that we have considered only data wherein we have nearly
simultaneous radio and X-ray spectra. The radio/X-ray
observations discussed by \citet{nowak:02a} all have a large degree of
overlap between the radio and X-ray observing times, and they show
very little long time scale ($>1000$\,s) variability
\citep{corbel:00a}.  Radio observations C \& D occur approximately a
half day after their corresponding \rxte\ observations, but, again,
neither the radio data nor the \rxte\ data show any evidence for long
term variability within the observations themselves.  Still, the lack
of strict simultaneity might be a relevant factor in some of the
results discussed below.

\begin{figure}
\epsscale{1}
\plotone{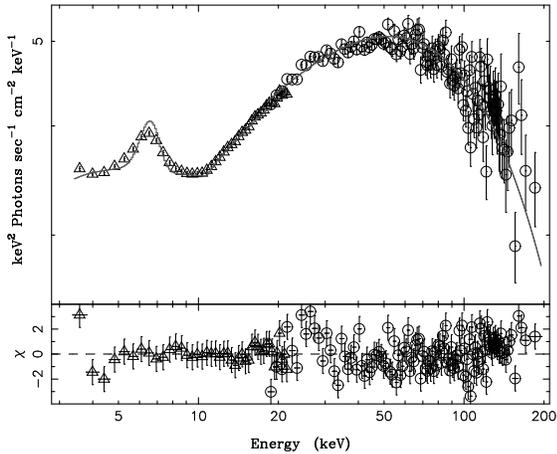}
\caption{Unfolded \rxte\ spectra of \cyg\ (Wilms et al., in prep.),
fit with an absorbed, exponentially cutoff, broken power-law
and a gaussian line. Residuals are from the proper forward folded
model fit. \label{fig:cyg}}
\end{figure}  

We further require that the source was bright enough to fit the
X-ray spectrum in both the \pca\ and the \hexte\ data.  As discussed
by \citet{corbel:03a} and \citet{wardzinski:02a}, at low flux levels
there is background contamination that affects the \rxte\ spectra of
\gx.  The spectra can be corrected adequately to yield reasonably
accurate integrated soft X-ray fluxes \citep{corbel:03a}; however, the
contamination is much more problematic for the spectral extrapolations
discussed here.

The observations were analyzed with the {\tt Interactive Spectral
Interpretation System} \citep[\isis;][]{houck:00a}.  For our purposes,
there are three major reasons for our use of \isis.  First, \isis\
uses {\tt s-lang} as a scripting language and hence has most of the
programmability of {\tt IDL (Interactive Data Language)} or {\tt
MATLAB}, while retaining all the models of {\tt XSPEC}
\citep{arnaud:96a}.  Second, data input without a response matrix
(i.e., the radio data) are automatically presumed to have an
associated diagonal response with one cm$^2$ effective area and one
second integration time.  Thus, we used a fairly straightforward {\tt
s-lang} script to convert the radio data from mJy to photon rate in
narrow bands around the observation frequencies. This then was used as
input for the simultaneous radio/X-ray fits. (So long as they are
narrow, the widths of the radio bands do not affect the fits.)  We
then set the fractional error bars of these `count rate' data equal to
the fractional error from the radio measurements.

The third reason for using \isis\ is that it treats `unfolded spectra'
(shown in Fig.~\ref{fig:spectra}) in a model-independent manner.  The
unfolded spectrum in an energy bin denoted by $h$, as used to create
Fig.~\ref{fig:spectra}, is defined by:
\begin{eqnarray}
F_{\rm unfold}(h) = \frac{ [ C(h) - B(h) ] / \Delta t }{\int R(h,E) A(E) dE} ~~,
\end{eqnarray}
where $C(h)$ is the total detected counts, $B(h)$ is the background
counts, $\Delta t$ is the integrated observation time, $R(h,E)$ is the
unit normalized response matrix describing the probability that a
photon of energy $E$ is detected in bin $h$, and $A(E)$ is the
detector effective area at energy $E$. Contrary to unfolded spectra
produced by {\tt XSPEC} (which are given by the model, rebinned to the
output energy bins of the response matrix, multiplied by the data
counts divided by the forward folded model counts, i.e., the ratio
residuals), this definition produces a spectrum that is independent of
the fitted model.  In Fig.~\ref{fig:spectra}, we over plot the fitted
model (using the internal resolution of the `ancillary response
function', i.e., the `arf', used in the fitting process). The plotted
residuals, however, are those obtained from a proper forward-folded
fit.

Although we previously have successfully fit the X-ray spectra with
sophisticated Comptonization models \citep{nowak:02a}, we obtain
surprisingly good fits for nine of the ten radio/X-ray spectra using
the following simple model (using the \isis/{\tt XSPEC} model
definitions): absorption (the {\tt phabs} model, with $N_{\rm H}$
fixed to $6\times10^{21}~{\rm cm}^2$) and a high energy, exponential
cutoff (the {\tt highecut} model) multiplying a doubly broken
power-law (the {\tt bkn2pow} model, with the first break being in the
far IR to optical regime, and the second break being constrained to
the 9--12\,keV regime) plus a gaussian line (with energy fixed at
6.4\,keV).  Results are presented in Table~2. When considering just
the X-ray spectra, a singly broken power-law fits all ten spectra,
with better results than any of the Comptonization models that we have
tried.  The 12--200\,keV power-law is typically seen to be harder than
the 3--12\,keV power-law by $\Delta \Gamma \approx -0.2$.  (In a
future work, correlations of this spectral break with overall hardness
will be presented for over 200 hard state spectra of \cyg; Wilms et
al., in prep.)

The phenomenological power-law model was of course chosen because
we are attempting to answer a phenomenological question: do the
extrapolated radio and X-ray spectra predict the amplitude of the IR
flux, and the location of any IR break?  The power-law provides the
simplest model to extrapolate.  The additional power-law break in the
9-12\,keV band is required by the data.  The energy band restriction
in the fitting process was to avoid spurious local minima, caused by
the break interfering with the gaussian line at low energy or with the
{\tt highecut} model at high energy.  All but one observation has
break energy values, including 90\% confidence level error bars, that
fall well within this 9--12\,keV range.

\begin{figure}
\epsscale{1}
\plotone{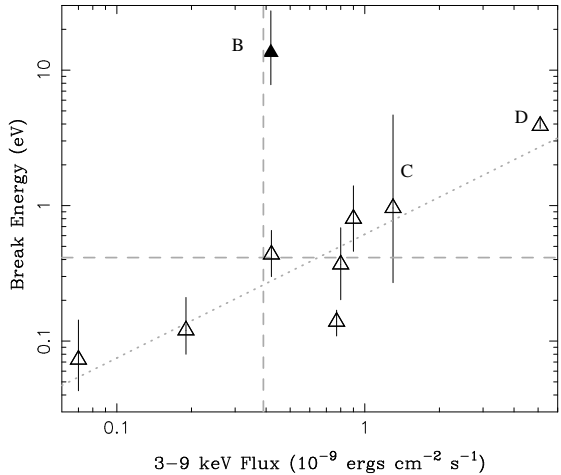}
\caption{Results of broken power-law fits to \gx, showing the location
of the break between the radio and soft X-ray power-law.  Labels refer
to Table 1, and are further described in the text. Here, and
throughout the remaining figures, solid points will refer to
observations A (not shown in the above ) and B, which have `optically
thin' radio spectra. Dashed lines show the approximate integrated
X-ray flux and IR spectral break energy previously observed in \gx\
(Corbel \& Fender 2002). The dotted line is $\propto \tfx^{0.91}$.
\label{fig:break}}
\end{figure}  

Phenomenologically speaking, the X-ray break is consistent with the
expectations from reflection models \citep{magdziarz:95a}.  However,
as shown in Fig.~\ref{fig:cyg}, the identical broken power-law model
described above provides a very good phenomenological description even
for \cyg\ hard state data with an extreme break.  Although not
mutually exclusive with the presence of reflection, such data require
at least two broad-band continuum components in the X-ray.  For
example, Fig.~\ref{fig:cyg} could be consistent with synchrotron and
synchrotron self-Compton from jet models \citep[][Markoff, Nowak, \&
Wilms, in prep.]{markoff:04a}, or a strong disk plus Comptonization
component from corona models.  These issues will be explored in
greater detail in a forthcoming work (Wilms et al., in prep.).

\subsection{Predicted Radio/X-ray Correlations}\label{sec:break}

Given only two radio points and an X-ray spectrum that is well-fit by
a singly broken power-law, it is not surprising that the doubly broken
power-law models work as well.  But how do the fitted locations of the
radio-to-soft X-ray break compare to the `IR coincidence' breaks, and
how do the fitted break locations scale with observed flux?  In
Fig.~\ref{fig:break} we show the fitted radio-to-X-ray break location
as a function of 3--9\,keV integrated flux.  We also show in this
figure the approximate integrated 3--9\,keV flux and the IR break
location for the 1997 observation discussed by \citet{corbel:02a}.
For our \gx\ observations of comparable 3--9\,keV flux, the doubly
broken power-law models do indeed produce a break in the IR.  Looking
over the whole span of observed integrated X-ray fluxes, however, we
see that the model fits presented here have predicted radio-to-X-ray
breaks ranging all the way from the far IR to the blue end of the
optical (and into the X-ray, if one also considers observation B
[40108-01-04], which has an `optically thin' radio spectrum).

\begin{figure}
\epsscale{1} \plotone{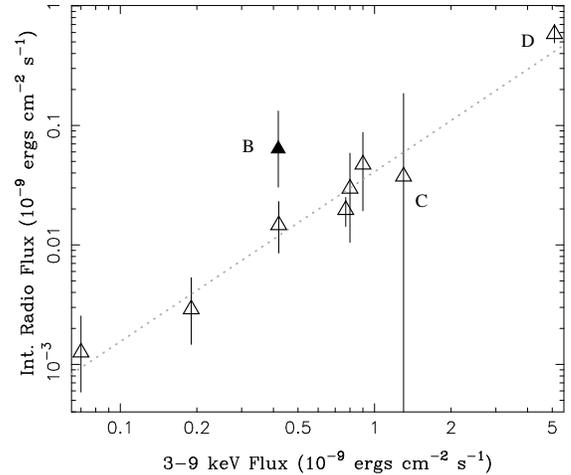}
\caption{The flux of the radio power-law, integrated from zero energy
to the spectral break between the radio and soft X-ray, vs. the
3--9\,keV (\pca) flux.  Labels refer to Table 1 (Observation A not
shown), and are further described in the text. The dotted line is
$\propto \tfx^{1.4}$.
\label{fig:fluxcor}}
\end{figure}

To assess the correlation of radio-to-X-ray break energy with
integrated X-ray flux, we exclude observation B (40108-01-04), which
has an optically thin radio spectrum, and perform a regression
analysis on the remaining eight data points.  We weight the data
uniformly, which is equivalent to assuming that intrinsic variations
in any correlation dominate over statistical errors.  (In the results
discussed below, the derived regression slopes and errors encompass the
values obtained if one weights the data by their error bars.)  We find
that the radio-to-X-ray break energy, in eV, scales with the 3--9\,keV
integrated flux as $0.61 \tfx^{0.91\pm0.18}$.  Does this agree with
models wherein the observed soft X-ray spectrum is the optically thin
synchrotron emission from a jet?

Using a scale invariance {\sl Ansatz} to describe the jet physics
\citep[][see also \citealp{falcke:95a} and
\citealp{markoff:03a}]{heinz:03a,heinz:04a}, we can predict the
scaling between the integrated X-ray synchrotron flux and the
radio-to-X-ray break frequency where the jet becomes optically thin to
synchrotron self-absorption.  For simplicity, we make the assumption
that, at the base of the jet, the fraction of particle pressure to
magnetic pressure is independent of accretion rate/jet power. 

Given the above assumptions, the optically thick radio flux follows
\citep{heinz:03a}:
\begin{equation}
  \cfr \propto B^{\frac{\scriptstyle [2p - (p+6)\alpha_{\rm r} +
  13]}{\scriptstyle [p+4]}} ~~,
\end{equation}
where $\alpha_{\rm r}$ is the radio spectral index, $p$ is the
power-law index of the electron spectrum (at energies below any
synchrotron cooling regime), and $B$ is the magnetic field at the base
of the jet.  From \citet{heinz:04a}, using the same assumptions, the
X-ray synchrotron flux integrated over a fixed energy band follows
\begin{equation}
  \tfx \propto B^{\scriptstyle [1 + 2p + 3\alpha_{\rm x}]} ~~,
\end{equation}
where $\alpha_{\rm x}$ is the X-ray spectral index. (We again are
using the convention that $\alpha \equiv 1 - \Gamma$.)  Combining
these expresions, we find
\begin{equation}
  \cfr \propto \tfx^{ \frac {\scriptstyle [2p - (p+6)\alpha_{\rm r} + 13]}
   {\scriptstyle [p+4][1 + 2p +3\alpha_{\rm x}] }}~~.
\end{equation}
If the X-ray band is unaffected by synchrotron cooling, the relation
follows $\cfr \propto \tfx^{0.69 \rightarrow 0.81}$, for values of
$2.6 < p < 2$, $-0.8 < \alpha_x = (1-p)/2 < -0.5$, and $0 <
\alpha_{\rm r} < 0.2$.  This scaling is consistent with previous
observations, and is also consistent with prior descriptions of the
jet model's predictions \citep[i.e.,][]{corbel:03a,markoff:03a}.

The lower break frequency, $\nu_{\rm b}$, from an optically thick to
optically thin spectrum is proportional to \citep{heinz:03a}
\begin{equation}
  \nu_{\rm b} \propto B^{\frac{\scriptstyle [6+p]}{\scriptstyle [4+p]}} ~~.
\end{equation}
We thus can write the dependence of $\nu_{\rm b}$ upon $\tfx$ as
\begin{equation}
  \nu_{\rm b} \propto \tfx^{ \frac{ \scriptstyle [6+p]}
  {\scriptstyle [1 + 2p + 3\alpha_{\rm x}][4+p]}} \propto 
  \tfx^{ \frac{ \scriptstyle 2 [p+6]}{\scriptstyle [p+4] [p+5] }} ~~,
\label{eq:break}
\end{equation}
where for the latter relation we have used $\alpha_{\rm x}=(1-p)/2$
from standard synchrotron theory.  Again taking $-0.8 < \alpha_{\rm x}
< -0.5$, we obtain $\nu_{\rm b} \propto F_{\rm x}^{0.34}$ to $\nu_{\rm
b} \propto F_{\rm x}^{0.38}$.  This prediction is flatter than the
observed dependence of extrapolated break frequency upon X-ray flux.

Eq.~(\ref{eq:break}) represents the scaling of the \emph{actual
location} of the break from an optically thick to optically thin
spectrum (as appropriate for the `IR coincidence' yielding the
appropriate location of an observed break).  Our observations instead
yield the scaling of the \emph{inferred location} of the break, which
has an additional dependence upon the scalings of the spectral slopes,
$\alpha_{\rm r}$ and ${\alpha_{\rm x}}$.  Specifically, if one assumes
that there is an underlying radio/X-ray correlation given by $\cfr
\propto \tfx^\beta$, yet allows evolution of the radio and X-ray
spectral slopes, one can show that for the inferred break, $\nu_{\rm b}$,
\begin{eqnarray}
\log \left ( \frac{\nu_{\rm b}}{\nu_{\rm 0}} \right ) 
~ = ~  
\left [ 
   \left ( \alpha_{\rm r} - \alpha_{\rm x} \right ) + 
   \Delta \alpha_{\rm r} - \Delta \alpha_{\rm x} 
\right ]^{-1}
~ \times ~
\cr 
\left [ 
   ( 1 - \beta ) \log  \left ( \frac{F_{\rm x}}{F_0} \right )  
   + \log \left ( \frac{\nu_{\rm r}}{\nu_0} \right ) \Delta \alpha_{\rm r}
   - \log \left ( \frac{\nu_{\rm x}}{\nu_0} \right ) \Delta \alpha_{\rm x} 
\right ] ~,
\end{eqnarray}
where $\nu_0$ is the break energy measured at an X-ray flux of $F_0$,
$\nu_{\rm r}$, and $\nu_{\rm x}$ are the frequencies at which the radio
and X-ray fluxes, respectively, are measured, and $\Delta \alpha_{\rm
r}$ and $\Delta \alpha_{\rm x}$ are the changes in radio and X-ray
spectral indices as the X-ray flux changes to $F_{\rm x}$. (Note that
small correction factors for the dependence of integrated X-ray flux
upon $\alpha_{\rm x}$, as opposed to solely power law normalization,
have been omitted.)

In the absence of evolution of the spectral indices, one expects the
break frequency to scale with X-ray flux with a power of
$(1-\beta)/(\alpha_{\rm r} - \alpha_{\rm x}) \approx 0.35$, as
described above.  However, as noted elsewhere
\citep{wilms:99aa,nowak:02a}, many of the observations of \gx\ show a
softening/hardening with increasing/decreasing X-ray flux.  (For
sources with optically thick radio spectra, notable exceptions to this
trend are Obs IDs 40108-02-03 and 70109-01-02.  A detailed discussion
of flux/hardness trends can be found in \citealt{nowak:02a}.)
Comparing Obs ID 40108-02-02 to 20181-01-01, $\Delta \alpha_{\rm x}
\approx 0.06$ for $\log(F_{\rm x}/F_0) \approx 0.7$.  Combining this
with $\alpha_{\rm r} + \alpha_{\rm x} \approx 0.8$ and $\log(\nu_{\rm
x}/\nu_0) \approx 4.9$ means that the break energy should scale with
with X-ray flux with a power of $(1 - \beta + 0.5)/0.8 \approx 1$.  This
is in much closer agreement with the observed trends.

Contrary to $\alpha_{\rm x}$, there are no obvious trends for a
dependence of $\alpha_{\rm r}$ upon X-ray flux.  (This will be
explored in a future work, with a larger set of observations from 
2002 and 2004 outbursts of \gx; Corbel et al., in prep.)  The mean value for all
the optically thick radio data is $\langle \alpha_{\rm r} \rangle =
0.16$, with a scatter of $\pm 0.05$.  Combining this with
$\log(\nu_{\rm r}/\nu_0) \approx -3.5$, and the fact that from the
faintest to brightest observation is a factor of $10^{1.9}$, we expect
the variations in radio slope to account for correlation slope
variations of up to $\pm 0.1$.

We thus find that the observed correlation of inferred break energy is
inconsistent with the expectations from the very simplest X-ray
synchrotron jet model (i.e., the jet spectra scaling with input power,
and no evolution of the radio or X-ray slopes).  The observed trends
can be recovered, however, if the differences from the simplest
expectations are primarily driven by evolution of the X-ray spectral
slope, $\alpha_{\rm x}$, with X-ray flux, with possible lesser
contributions from variations of the radio spectral slope,
$\alpha_{\rm r}$.  We shall discuss this point further in
\S\ref{sec:sum}.

\begin{figure}
\epsscale{1} \plotone{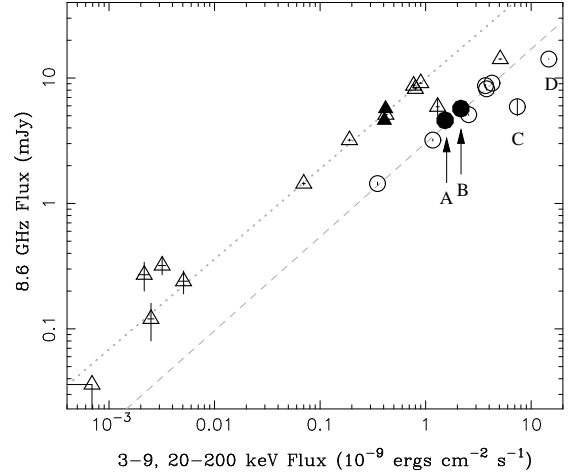}
\caption{3--9\,keV (triangles) and 20--200\,keV (circles) X-ray flux
(units of $10^{-9}~{\rm ergs~cm~s^{-1}}$) from pointed \rxte\
observations of \gx\ vs. 8.6 GHz radio flux (mJy).  Labels refer to
Table 1, and are further described in the text.  Lines show power-laws
of the form $\tfx^b$, with $b = 0.72$ (dotted line) and $b = 0.75$
(dashed line).
\label{fig:gxcor}}
\end{figure}

A further interesting correlation is seen when one looks at the
dependence of radio flux, integrated between zero energy and the break
frequency, upon the integrated X-ray flux, as we show in
Fig.~\ref{fig:fluxcor}.  Considering both in units of $10^{-9}~{\rm
erg~cm^{-2}~s^{-1}}$, and again weighting the data uniformly, we
find $\tfr = 0.04 \tfx^{1.42\pm0.10}$.  This implies that, even
ignoring the energy of the particles and fields in the jet, the `jet
radiation' up to the inferred break is 3--10\% of the integrated
3--9\,keV X-ray flux, and is 0.5--5\% of the 3--200\,keV flux.

Again, we can compare this result to expectations from the
simplest jet models. Utilizing our previous estimates, we find for the
product of $\cfr \nu_{\rm b}$:
\begin{equation}
  \cfr \nu_{\rm b} \propto F_{\rm x}^{\frac{ \scriptstyle 
   [ 19+3p - \alpha_{\rm r} (p+6) ]}{\scriptstyle [ p+4 ][ 1 + 2p +
  3\alpha_{\rm x} ]}}
\end{equation}
Again using the range of $-0.8 < \alpha_{\rm x} < -0.5$ and $0 <
\alpha_{\rm r} < -0.2$, this product should follow $\cfr \nu_{\rm b}
\propto \tfx^{1\rightarrow1.19}$. As before, for the
\emph{inferred value} we might expect to increase the scaling exponent
by $0.5 \pm 0.1$, if we allow for an evolution of $\alpha_{\rm r}$ and
$\alpha_{\rm x}$ with X-ray flux.

\begin{figure*}
\epsscale{1}  
\plottwo{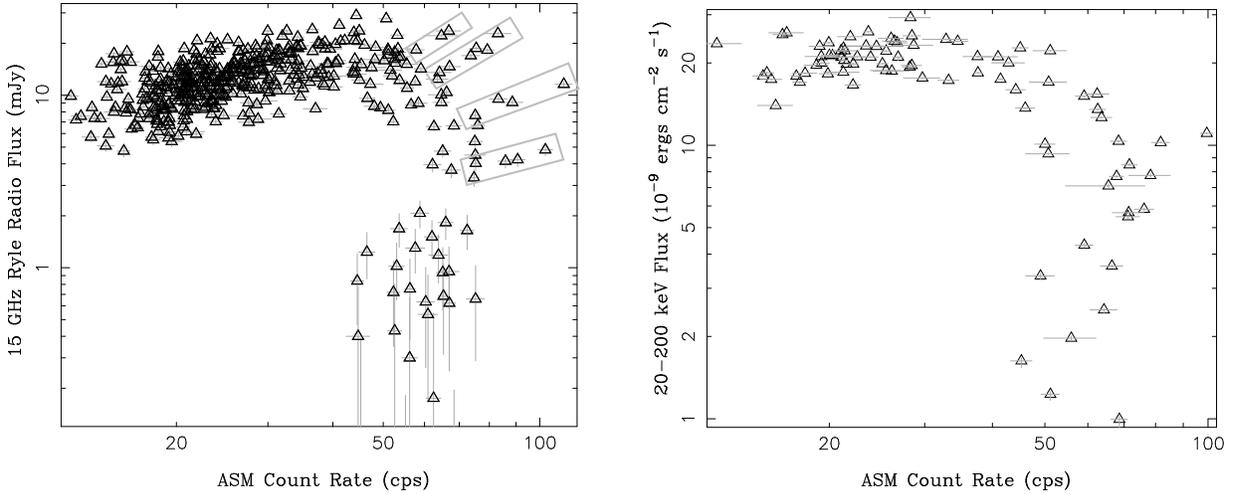}{cyg_asm_vs_hxt.ps}
\caption{Left: 15 GHz Ryle radio flux (mJy) vs. \cyg\ daily mean \asm\
count rate (see text). Grey boxes highlight data associated with
failed state transitions, and the beginning of a transition to a
prolonged soft state. Right: 20--200\,keV flux (units of
$10^{-9}~{\rm ergs~cm~s^{-1}}$) from pointed \rxte\ observations.
vs. \cyg\ daily mean \asm\ count rate (see text).
\label{fig:asm}}
\end{figure*}

\section{Radio/X-ray Correlations in \gx}\label{sec:gxcor}

One of the interesting aspects of the radio/X-ray flux correlation in
\gx\ as identified by \citet{corbel:03a} is that it appeared to be
consistent\footnote{There are two radio points, however, from the 1999
decline into quiescence that deviate from the correlation
(Fig.~\ref{fig:gxcor}). Given the large degree of variability on week
time scales in the quiescent state \citep{kong:00a}, and the fact that
the magnitude of any time delay between the radio and X-ray flux
variations is unknown, it was not clear how significant these
deviations should be considered.} in both slope and amplitude between
the 1997 and 1999 low/hard states, despite the intervening radio
quiet, soft state outburst (see Fig.~\ref{fig:gxcor}).  The high
luminosity observations C and D [70109-01-02 and 40031-03-01],
however, were not part of the original correlation presented by
\citet{corbel:03a}.  These two observations occurred in 2002, 14 days
apart, in a bright hard state as \gx\ was rising into an even brighter
soft state.

In Fig.~\ref{fig:gxcor} we plot the 8.6\,GHz radio flux density\footnote{Note
that for observation C (70109-01-02), we have extrapolated the 2.4\,GHz
radio flux to 8.64\,GHz using the fitted power-law.}
vs. the 3--9 and 20--200\,keV X-ray fluxes.  As in \citet{corbel:03a},
we have performed a regression fit for all the 1997 and 1999
observations for which we could determine \emph{both} a 3--9\,keV
(\pca) and 20--200\,keV (\hexte) flux.  (However, we also plot the 1999 low
flux observations from \citealp{corbel:03a} that have only \pca\
data.) For the soft X-rays we find $\cfr = 9.9 \tfx^{0.73\pm0.02}$,
and for the hard X-rays we find $\cfr = 3.0 \tfx^{0.75\pm0.02}$.  The
slight differences between these correlation slopes and those
presented by \citet{corbel:03a} are due to our use of better
calibrated \rxte\ response matrices (especially for the \pca).  As
discussed by \citet{corbel:03a}, the correlations apply equally well
to the 1997 and the 1999 data, including the low flux 1999 data that
were not formally part of the regression fit.  The slopes between the
soft and hard X-rays are also reasonably consistent with one another
\citep{corbel:03a}.

The 2002 data, although only consisting of two points, deviate from
these previously viewed trends.  For these data, the radio flux
vs. the 3--9\,keV flux correlation appears to have the expected slope,
but more than a factor of two times lower amplitude.  The radio/hard
X-ray correlation, however, appears to be less reduced in amplitude,
but has a steeper slope, with $\cfr \propto \tfx^{1.3}$.  There
is one major uncertainty in these results: the radio data associated
with observations C and D were not strictly simultaneous with the
X-ray data.  We note, however, that the radio data, which falls below
the previously observed trends, was taken after the X-ray data while
the source was rising in flux on several day time scales.

It is also possible that observations C and D, rather than being
on a lower radio amplitude parallel track, are instead indicating a
turnover in the correlation occuring at high X-ray flux.  Such a
turnover has been observed for the correlation seen in \cyg\
\citep[][and Fig.~\ref{fig:asm}]{gallo:03a}.  If this latter
hypothesis is correct, then given the observed 14.2\,mJy flux for the
highest flux observation, D, the deficit is a factor of 2.3 for the
3--9\,keV correlation (which predicts 32.4\,mJy), but only a factor of
1.6 for the 20--200\,keV flux correlation (which predicts 22.7\,mJy).
Extrapolating the observed 2.4\,GHz radio flux for observation C
to 8.64\,GHz yields $5.9\pm0.9$\,mJy.  This is a factor of
$2^{+0.4}_{-0.2}$ below the value predicted by the radio/3--9\,keV
flux correlation, and a factor of $2.2^{+0.4}_{-0.3}$ below the value
predicted by the radio/20--200\,keV flux correlation.

A further interesting fact to note also comes from the radio/hard
X-ray correlation.  Observations A and B (40108-01-03 and 40108-01-04)
were the first observations taken after the 1999 return to the
radio-loud hard state, and both have optically thin (i.e.,
$\alpha_{\rm r} < 0$) radio spectra.  Although they have nearly
identical 3--9\,keV integrated fluxes, they have different radio
fluxes and they straddle the radio/soft X-ray flux correlation shown
in Fig.~\ref{fig:gxcor}.  However, these points actually lie along the
radio/hard X-ray flux correlation, as they have significantly
different 20--200\,keV fluxes.  This might argue that the radio/hard
X-ray flux is the more fundamental relationship.

\section{Radio/X-ray Correlations in Cyg~X-1}\label{sec:cygcor}

The radio/X-ray correlations in \gx\ present us with two, not
necessarily mutually exclusive, possibilities.  Either the radio/X-ray
correlation is more fundamentally tied to the hard X-rays, and/or
different instances of the hard state can present the correlation with
different amplitudes. To further explore these possibilities, we turn
to radio/X-ray observations of \cyg\
\citep{pottschmidt:02a,gallo:03a}. As described in
\citet{pottschmidt:02a}, the radio data are 15\,GHz observations
performed at the Ryle Telescope, Cambridge (UK).  These are single
channel observations, so a radio spectral slope cannot be determined.
Most of these observations have occurred simultaneously with pointed
\rxte\ observations \citep{pottschmidt:02a}, and nearly all have very
good contemporaneous coverage by the \rxte\ \textsl{All Sky Monitor}
(\asm). The \asm\ provides coverage in the 1.5--12\,keV energy band
\citep{remillard:97a}, and therefore allows us to assess the
radio/soft X-ray correlation in \cyg.

In Fig.~\ref{fig:asm} we plot the daily average \asm\ count rate
vs. the daily average 15\,GHz flux.  Here we only include days for
which there are at least 25 \asm\ data points (representing 70 second
dwells) with a reduced $\chi^2<1.5$ for the \asm\ solution
\citep[see][]{remillard:97a}.  Ranging from approximately 10--50\,cps
in the \asm\, there is a clear log-linear correlation between the
radio flux and the \asm\ count rate.  As for \gx, the radio flux
rises more slowly than the \asm\ count rate ($\cfr$ scales
approximately as the 0.8 power of the \asm\ count rate). \cyg,
however, shows much more scatter in the amplitude of the correlation
than does \gx. This scatter is \emph{not} dominated by variations
related to the orbital phase of the binary system
\citep[cf.,][]{brocksopp:99b}.

As noted by \citet{gallo:03a}, there is a sharp roll-over for higher
\asm\ count rates.  However, one can clearly discern on the shoulder
of this roll-over (i.e., the upper right corner of Fig.~\ref{fig:asm})
four `spokes', consisting of 2--5 data points each.  In these spokes,
the radio/X-ray correlation appears to hold to high count rates.  We
have confirmed\footnote{We have used the `vwhere' program
\citep{noble:05a} which allows one to interactively filter one set of
data, e.g., a color-color or color-intensity diagram, and then apply
those filters to the same observations visualized in a different way,
e.g., a time-intensity diagram.  (The program is available at {\tt
http://space.mit.edu/cxc/software/slang/modules/}) } that each of these times
are associated with `failed transitions' to the soft state, as
described by \citet{pottschmidt:02a}, except for the lowest amplitude
of these spokes, which occurs immediately preceding a prolonged soft
state outburst.  In many ways these failed state transitions and the
early stage of a soft state outburst for \cyg\ are reminiscent of the
data presented here for the 2002 \gx\ hard state.  As for those latter
\gx\ data, the \cyg\ spectra remain hard, although not as hard as
low-luminosity hard states, to very high flux levels.

\begin{figure}
\epsscale{1}  
\plotone{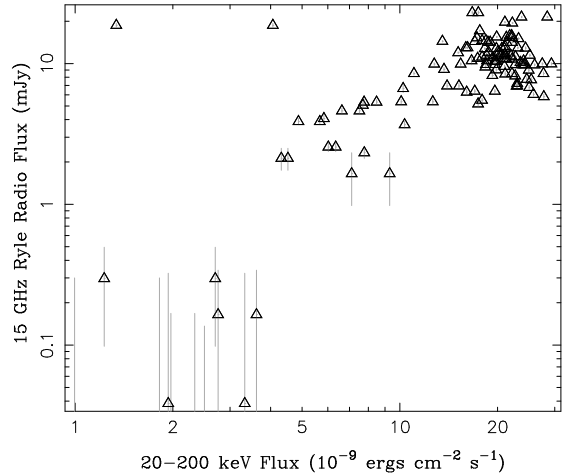}
\caption{20--200\,keV flux (units of $10^{-9}~{\rm ergs~cm~s^{-1}}$)
vs. the daily average 15\,GHz Ryle radio flux (mJy) for pointed observations of
\cyg. \label{fig:cygcor}}
\end{figure}

In Fig.~\ref{fig:asm} we also plot the daily average \asm\ count rate
vs. the 20--200\,keV flux from our pointed \rxte\ observations taken
during the same 24 hour period \citep{pottschmidt:02a}.  Here,
however, we only require 3 \asm\ data points per average, for a
resulting 85 observations.  We see that hard X-ray/\asm\ correlation
traces a similar pattern to the radio/\asm\ correlation.  Indeed, when
we plot the hard X-ray flux vs. the daily average radio flux
(representing 129 separate hard X-ray pointed observations), we obtain
a log-linear relationship, as shown in Fig.~\ref{fig:cygcor}
\citep[see also][]{gleissner:04b}.  (Two high radio flux, but low
X-ray flux, data points are seen in the upper left of this figure;
these represent intra-day hard X-ray variations.)  Returning to the
question that we posed in \S\ref{sec:cygcor} of whether we were seeing
evidence for parallel tracks in the radio/X-ray correlation, or
whether we were seeing the correlation being more fundamental to the
hard X-ray band, in \cyg\ the answer seems to be that both effects are
discernible, but fundamentally the radio flux density does appear to
be tied to the hard X-ray emission.

\section{Summary}\label{sec:sum}

In this work we have considered ten simultaneous or near
simultaneous \rxte/radio observations of \gx, and over one hundred
\rxte/radio observations of \cyg.  We have fit the former spectra with
a very simple, but remarkably successful, phenomenological model
consisting of a doubly broken power-law with an exponential roll-over
plus a gaussian line.  For \gx, the break between the radio and soft
X-ray power-laws occurs in the IR to optical range, in agreement with
the prior work of \citet{motch:85a} and \citet{corbel:02a} (i.e., the
`IR coincidence').  In contrast to these prior works, we have fit the
X-ray data in `detector space' and provided a quantitative assessment
of the extrapolated break location.

Is the `IR coincidence' just a coincidence? The spectral fits to the
\gx\ data suggest that the answer is `possibly not'.  All of the
data with optically thick radio spectra are extremely well-fit by
doubly broken power-laws, with the break being in the IR regime.
Although the scaling of this break frequency with X-ray flux does not
agree with the predictions of the simplest X-ray synchrotron jet
models, if one allows for softening of the X-ray spectrum with
increasing X-ray flux, then the jet model predictions agree with the
observed correlations.  There are several possibilities why such a
softening could occur.  In terms of the jet model, for example, there
could be cooling of the electron spectrum leading to a steepening in
the X-rays.  A second possibility is that the soft X-rays are being
contaminated by excess emission from a disk component that becomes
more prominent with increasing flux.  What is clear is that in order
for jet models to explain both the scaling of $\cfr \propto
\tfr^{0.72}$ and $\nu_b \propto \tfr^{0.92}$, they must be more
sophisticated than the very simplest considerations.  Fits to some of
these data sets with such models, which include disk emission,
synchrotron self-Compton emission, etc., are currently being
considered in detail, and will be discussed in a future work (Markoff,
Nowak, \& Wilms, in prep.).  The results presented here, however, are
consistent with the notion that at least some fraction of the
observed soft X-rays may be attributable to emission from the jet, as
opposed to disk or corona.

On the other hand, other facts suggest that the answer to the question
of the `IR coincidence' being just that is `maybe'.  We have tentative
evidence in \gx, and firmer evidence in the \cyg\ failed state
transitions and soft state transition, that the correlation between
radio flux and integrated X-ray flux can take on different amplitudes
during different hard state episodes.  There is also evidence in \cyg\
that the radio/X-ray correlation is more fundamental to the hard X-ray
band.  In jet models, this band, which essentially encompasses the
third, highest energy, power-law component in our model fits (and also
encompasses the exponential cutoff), is possibly attributable to the
synchrotron self-Compton (SSC) emission from the base of the jet
\citep[see, e.g.,][]{markoff:03a,markoff:04a}.  It is therefore quite
reasonable to expect a strong coupling between the radio and hard
X-ray flux; however, these models are more complex than simple pure
synchrotron models, and are only now beginning to be explored
quantitatively \citep{markoff:03a,markoff:04a}.  It is also worth noting that such
radio/X-ray couplings can be expected in Compton corona models
\cite[e.g.,][]{meier:01a,merloni:02a}.

More problematic for pure jet models are \gx\ observations A and B
(40108-01-03, 40108-01-04), both of which occurred very shortly after
the 1999 return to the hard state.  They have `typical' hard state
properties in terms of spectra and variability \citep{nowak:02a}, but
have optically thin radio spectra that do not neatly extrapolate into
the observed soft X-ray spectra.  They satisfy the radio
flux density/integrated X-ray flux correlation; however, they agree better
with the radio/hard X-ray flux correlation, for which these two
observations are cleanly differentiated from one another
(Fig.~\ref{fig:gxcor}).  It is possible that they represent an early,
transient phase of the jet as it forms.  It then remains a
theoretical question for the jet model whether it can describe a
scenario where the basic radio/X-ray flux correlation and `typical'
steady-state X-ray spectra hold, while the radio has not yet settled
into a steady `optically thick' state.

The results presented here suggest, at the very least, some obvious
observational strategies to arrive at a more definitive answer to the
question that we have posed.  Given the break energy correlations, it
would be extremely useful to have not only a radio amplitude for each
X-ray observation, but also a radio slope.  Furthermore, the inferred
break for the brightest observation, D (40031-03-01), occurs in the
blue end of the optical.  Thus, ideally multi-wavelength observations
would consist of radio, broad band X-ray, and IR through optical
coverage \citep[see, for example,][which suggest the possibility of
jet synchrotron radiation extending into the optical regime for the
hard state of XTE J1118+48]{hynes:03b,malzac:04a}.  This is an
admittedly difficult task, but BHC are demonstrating via spectral
correlations that all these energy regimes are fundamentally related
to activity near the central engine, i.e., the black hole, of the
system.

Finally, it is important to obtain multi-wavelength observations of
multiple episodes of each of the spectral states.  For example, if
there are indeed `parallel tracks' in the radio/X-ray correlations, it
would be interesting to determine whether the amplitude of the
radio/X-ray correlation is related to the flux at which the
outbursting source transits from the low/hard to high/soft state.  We
note that the low amplitude radio/X-ray correlation of the 2002 \gx\
outburst was associated with a very high flux level for the
hard-to-soft state transition.  The low amplitude `spokes' in the
\cyg\ radio/X-ray correlation might be similar in that they are
associated with failed state transitions.  If such observations can be
made with more quantitative detail, we will have vital clues to
determining the relative contributions of coronae and jets, and the
coupling between these two components, for black hole binary systems.

\acknowledgments It is a pleasure to acknowledge useful conversations
with Sera Markoff and Jeroen Homan, and for pointing us towards the
2002 data used in this work.  A debt of gratitude also goes to Mike
Noble and John Houck for invaluable software assistance.  This work
has been supported by NASA grant SV3-73016 and NSF grant INT-0233441.


\begin{deluxetable*}{cccccccccccc}
\tablefontsize{\scriptsize}
\tablewidth{0pt}
\tablecaption{\gx\ Observed Fluxes (1$\sigma$ confidence).}
\tablehead{ \colhead{Obs ID} &  & \colhead{y.m.d} & \colhead{0.84\,GHz} 
            & \colhead{1.38\,GHz} & \colhead{2.40\,GHz} & \colhead{4.80\,GHz}  
            & \colhead{8.64\,GHz} & \colhead{3--9\,keV} & \colhead{9--20\,keV} 
            & \colhead{20--100\,keV} & \colhead{100--200\,keV} \\
            & & & \multicolumn{5}{c}{(mJy)} 
            & \multicolumn{4}{c}{(${\rm 10^{-9}~ergs~cm^{-2}~s^{-1}}$)} }
\startdata
20181-01-01 & & 1997.02.03
            & \errpm{7.0}{0.6} & \nodata  & \nodata & \nodata  
            & \errpm{9.1}{0.1}
            & 0.90 & 0.92 & 2.98 & 1.27
            \\
\tabspace
20181-01-02 & & 1997.02.10
            & \errpm{5.5}{0.7} & \nodata  & \nodata & \nodata  
            & \errpm{8.2}{0.2}
            & 0.80 & 0.82 & 2.67 & 1.11
            \\
\tabspace
20181-01-03 & & 1997.02.17
            & \errpm{5.6}{0.7} & \errpm{5.4}{0.2}  & \errpm{6.0}{0.2} & \nodata  
            & \errpm{8.7}{0.2}
            & 0.77 & 0.78 & 2.63 & 1.06
            \\
\tabspace
40108-01-03 & A & 1999.02.12 & \nodata  & \nodata & \nodata  
            & \errpm{6.34}{0.08} & \errpm{4.60}{0.08}
            & 0.41 & 0.39 & 1.18 & 0.48
            \\
\tabspace
40108-01-04 & B & 1999.03.03 & \nodata  & \nodata & \nodata  
            & \errpm{6.07}{0.06} & \errpm{5.74}{0.06}
            & 0.42 & 0.42 & 1.51 & 0.66
            \\
\tabspace
40108-02-01 & & 1999.04.02 & \nodata & \nodata & \nodata
            & \errpm{4.75}{0.06} & \errpm{5.10}{0.06}
            & 0.42 & 0.44 & 1.60 & 0.97
            \\
\tabspace
40108-02-02 & & 1999.04.22 & \nodata & \nodata & \nodata
            & \errpm{2.92}{0.06} & \errpm{3.20}{0.06}
            & 0.19 & 0.21 & 0.74 & 0.43
            \\
\tabspace
40108-02-03 & & 1999.05.14 & \nodata & \nodata & \nodata
            & \errpm{1.25}{0.06} & \errpm{1.44}{0.06}
            & 0.07 & 0.07 & 0.21 & 0.14
            \\
\tabspace
70109-01-02 & C & 2002.04.03 & \nodata & \errpm{4.83}{0.20} 
            & \errpm{5.13}{0.11}
            & \nodata & \nodata
            & 1.30 & 1.47 & 5.34 & 2.09
            \\
\tabspace
40031-03-01 & D & 2002.04.18 & \nodata & \nodata & \nodata 
            & \errpm{13.04}{0.06} & \errpm{14.16}{0.07}
            & 5.10 & 4.97 & 12.3 & 2.42
            \\
\enddata

\tablecomments{X-ray flux errors are taken to be 3.5\%, the
average of the fitted normalization difference between \pca\ and
\hexte, since this exceeds the statistical uncertainties.}

\end{deluxetable*}

\begin{deluxetable*}{llccccccccccr}
\tablefontsize{\scriptsize}
\tablewidth{0pt}
\tablecaption{\gx\ Spectral Fits (90\% Confidence Level 
       errors).}
\tablehead{ \colhead{Obs ID} &  & \colhead{${\rm E_{cut}}$} & 
     \colhead{${\rm E_{fold}}$} & \colhead{$\alpha_{\rm r}$}
     & \colhead{${\rm E_{b-r}}$} & \colhead{$\Gamma_{\rm soft}$}
     & \colhead{${\rm E_{b-x}}$}  & \colhead{$\Gamma_{\rm hard}$} & \colhead{${\rm A_{line}}$}
     & \colhead{$\sigma_{\rm line}$} & \colhead{Con.}
     & \colhead{$\chi^2$/DoF} \\
     & & \colhead{(keV)} & \colhead{(keV)}
     & & \colhead{(eV)} & & \colhead{(keV)} & & ($10^{-3 }$ ) & (keV) }
\startdata
20181-01-01 &
            & \errtwo{37}{7}{6} & \errtwo{155}{22}{19}
            & \errtwo{0.11}{0.08}{0.07}
            & \errtwo{0.80}{1.22}{0.48} & \errtwo{1.69}{0.01}{0.01} 
            & \errtwo{11.3}{0.6}{0.5} & \errtwo{1.49}{0.03}{0.02}
            & \errtwo{2.1}{0.2}{0.3} & \errtwo{0.6}{0.1}{0.1} 
            & \errtwo{0.97}{0.01}{0.02} & 110/108 \\
\tabspace
20181-01-02 &
            & \errtwo{42}{5}{5} & \errtwo{140}{16}{14}
            & \errtwo{0.17}{0.10}{0.09}
            & \errtwo{0.37}{0.66}{0.23} & \errtwo{1.68}{0.01}{0.00} 
            & \errtwo{11.0}{0.6}{0.5} & \errtwo{1.48}{0.02}{0.02} 
            & \errtwo{1.8}{0.3}{0.2} & \errtwo{0.7}{0.1}{0.2} 
            & \errtwo{0.95}{0.01}{0.01} & 162/114 \\
\tabspace
20181-01-03 &
            & \errtwo{53}{14}{8} & \errtwo{133}{33}{31}
            & \errtwo{0.26}{0.04}{0.04}
            & \errtwo{1.39}{0.61}{0.40} & \errtwo{1.68}{0.01}{0.00} 
            & \errtwo{10.7}{0.6}{0.7} & \errtwo{1.51}{0.02}{0.02}
            & \errtwo{1.7}{0.2}{0.2} & \errtwo{0.6}{0.1}{0.1} 
            & \errtwo{0.98}{0.02}{0.02} & 90/90 \\
\tabspace
40108-01-03 & A 
            & \errtwo{44}{9}{11} & \errtwo{165}{58}{41}
            & \nodata
            & \nodata & \errtwo{1.80}{0.01}{0.01} 
            & \errtwo{10.2}{1.0}{0.5} & \errtwo{1.59}{0.02}{0.03}
            & \errtwo{1.2}{0.1}{0.2} & \errtwo{0.7}{0.1}{0.1} 
            & \errtwo{1.03}{0.01}{0.01} & 64/77 \\
\tabspace
40108-01-04 & B 
            & \errtwo{69}{14}{13} & \errtwo{138}{59}{48}
            & \errtwo{-0.09}{0.04}{0.05}
            & \errtwo{13.5}{37.}{8.0} & \errtwo{1.69}{0.01}{0.01} 
            & \errtwo{10.6}{0.6}{0.7} & \errtwo{1.53}{0.01}{0.02}
            & \errtwo{0.9}{0.1}{0.1} & \errtwo{0.6}{0.1}{0.2} 
            & \errtwo{1.03}{0.02}{0.02} & 106/103 \\
\tabspace
40108-02-01 & 
            & \errtwo{32}{21}{8} & \errtwo{345}{133}{79}
            & \errtwo{0.12}{0.06}{0.05}
            & \errtwo{0.44}{0.43}{0.20} & \errtwo{1.67}{0.01}{0.01} 
            & \errtwo{10.5}{0.6}{0.6} & \errtwo{1.45}{0.03}{0.02}
            & \errtwo{0.7}{0.2}{0.1} & \errtwo{0.5}{0.2}{0.2} 
            & \errtwo{0.97}{0.02}{0.02} & 115/93 \\
\tabspace
40108-02-02 &  
            & \errtwo{44}{33}{18} & \errtwo{304}{322}{142}
            & \errtwo{0.16}{0.07}{0.08}
            & \errtwo{0.12}{0.18}{0.06} & \errtwo{1.63}{0.01}{0.01} 
            & \errtwo{10.7}{1.2}{1.4} & \errtwo{1.47}{0.03}{0.05}
            & \errtwo{0.4}{0.1}{0.1} & \errtwo{0.5}{0.1}{0.2} 
            & \errtwo{0.95}{0.03}{0.03} & 83/67 \\
\tabspace
40108-02-03 & 
            & \nodata & \nodata
            & \errtwo{0.24}{0.12}{0.12}
            & \errtwo{0.07}{0.16}{0.04} & \errtwo{1.68}{0.01}{0.02} 
            & \errtwo{9.0}{3.}{0.0} & \errtwo{1.62}{0.03}{0.05}
            & \errtwo{0.1}{0.1}{0.0} & \errtwo{0.6}{0.2}{0.3} 
            & \errtwo{0.95}{0.07}{0.08} & 26/44 \\
\tabspace
70109-01-02 &  C
            & \errtwo{34}{2}{2} & \errtwo{112}{6}{5}
            & \errtwo{0.11}{0.14}{0.11}
            & \errtwo{0.95}{14.3}{0.81} & \errtwo{1.61}{0.01}{0.01} 
            & \errtwo{10.3}{0.4}{0.4} & \errtwo{1.37}{0.01}{0.02}
            & \errtwo{2.2}{0.3}{0.4} & \errtwo{0.4}{0.1}{0.1} 
            & \errtwo{0.99}{0.01}{0.01} & 180/148 \\
\tabspace
40031-03-01 & D 
            & \errtwo{26}{1}{2} & \errtwo{69}{2}{3}
            & \errtwo{0.14}{0.02}{0.01}
            & \errtwo{3.88}{0.69}{0.41} & \errtwo{1.73}{0.01}{0.01} 
            & \errtwo{11.1}{0.5}{0.5} & \errtwo{1.53}{0.02}{0.03}
            & \errtwo{17.7}{2.0}{1.8} & \errtwo{0.8}{0.1}{0.1} 
            & \errtwo{0.99}{0.02}{0.01} & 143/130 \\
\enddata

\tablecomments{20181 data are from the 1997 hard state (Wilms et
al. 1999; Nowak, Wilms, \& Dove 2002), 40108 data are from the 1999
hard state (post-1998 soft state; Nowak, Wilms, \& Dove 2002), 70109
and 40031 data are from the 2002 very luminous hard state (post a 3
year duration quiescent state; Homan et al. 2004). Parameters in
italics were held fixed at that value for the fits.  Note that the
radio spectral index, $\alpha_{\rm r}$, is defined to be positive for
`optically thick' spectra, and is related to the usual X-ray astronomy
definition of the photon index by $\alpha_{\rm r} = 1 -
\Gamma_{\rm r}$.}

\end{deluxetable*}

\end{document}